\documentclass[superscriptaddress,showpacs,twocolumn,amsmath,amssymb,floatfix,prl,aps]{revtex4}
\usepackage{graphicx}

\def\grad{\nabla_{\scriptscriptstyle\perp}}

\begin{document}

\title{Bogomol'nyi, Prasad and Sommerfield configurations in smectics}

\author{C.D. Santangelo}
\affiliation{Department of Physics, University of California, Santa Barbara, California  93106, USA}
\author{Randall D. Kamien}
\affiliation{Department of Physics and Astronomy, University of Pennsylvania, Philadelphia, PA 19104-6396, USA}

\date{\today}

\begin{abstract}
It is typical in smectic liquid crystals to describe elastic deformations with a linear theory when the elastic strain is small.  In smectics, certain essential nonlinearities arise from the requirement of rotational invariance.  By employing the Bogomol'nyi, Prasad and Sommerfield (BPS) decomposition and relying on boundary conditions and geometric invariants, we have found a large class of exact solutions. We introduce an approximation for the deformation profile far from a spherical inclusion and find an enhanced attractive interaction at long distances due to the nonlinear elasticity, confirmed by numerical minimization.
\end{abstract}


\maketitle

Exact solutions to nonlinear model problems offer deep insight into systems not at first apparent from simple linear approximations.   Liquid crystals with smectic-$A$ order, consisting of a stack of regularly spaced fluid layers, have particularly interesting nonlinear elastic properties that arise from considerations of rotational invariance.  The combination of these nonlinearities and thermal fluctuations is the source of anomalous elasticity in smectics~\cite{grinpelc}.  More recently, there has been some progress in the nonlinear theory of screw dislocations~\cite{KamBlue} and in the structure of isolated edge dislocations~\cite{edgedislocationtheory}.  Nonetheless, the consequence of these essential nonlinearities has been mostly ignored in the study of deformations induced by defects such as edge dislocations and inclusions embedded between the layers~\cite{sens1,sens2,sens3,clayparticles}.  In this letter we show that a large class of defect configurations can be probed by appealing to their implied boundary conditions and geometrical invariants.  We find, in particular, that though the layer displacements (shown in Fig. 1) can greatly differ between the linear and nonlinear theories, that the interactions between
edge dislocations is unchanged.

Recently, Brener and Marchenko have developed a solution for the deformation around a single edge dislocation that takes into account some of these nonlinearities~\cite{edgedislocationtheory}, and that differs significantly from the deformation profile derived from linear elasticity, even very far from the defect where the elastic strain and layer curvature is small.  This deformation profile has been confirmed in an experiment by Lavrentovich and Ishikawa~\cite{edgedislocationexpt} in a cholesteric finger texture.  Despite this success, it is difficult to understand how to use the solution to gain insight on the effects of nonlinearities on more complicated deformations (for example, curved or multiple edge dislocations).

\begin{figure}
	\resizebox{3.2in}{!}{\includegraphics{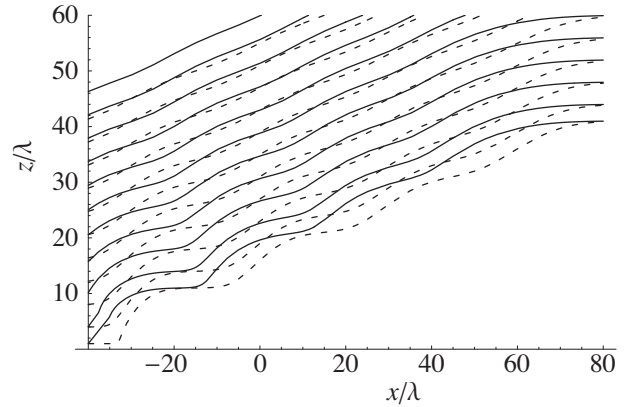}}
	\caption{\label{fig:layers} The layer displacement of four equally spaced, parallel edge dislocations in the nonlinear (solid) and linear (dashed) elasticity theories.  The Burgers vector for all dislocations is $20 \lambda$ and the layers shown are at $z=\lambda$, $z=4\lambda,8\lambda,\cdots$.  Note that the linear and nonlinear solutions differ by as much as $4\lambda$ and that the nonlinear solution relaxes more rapidly.}
\end{figure}

We have developed a large class of one-dimensional exact solutions which include the solution of ~\cite{edgedislocationtheory} as a limiting case.   This larger class includes, among others, certain multiple edge dislocation configurations.  The method of solution takes advantage of the fact that the free energy density for any one-dimensional deformation can be written as the sum of a perfect square and a total derivative.  This form also arises in the study of vortices in superconductors and kinks in nonlinear field theories~\cite{BPS}.  Expressing the free energy in this form reveals that the total deformation energy is always bounded from below (the Bogomol'nyi bound) and that so-called BPS configurations saturate this bound.  These BPS configurations satisfy a nonlinear differential equation of reduced order, which can be transformed into a linear equation through the Hopf-Cole transformation.  Thus, we are able to calculate the layer deformation and energy of these edge dislocation configurations exactly and ponder the effects of superposition.

This same technique produces approximate solutions to deformations that are not one-dimensional and that, in many cases, have lower energy than the deformation profile derived from linear elasticity.  In particular, we study the deformation profile far from an embedded inclusion in a lamellar phase, and the asymptotic interaction energy of two such inclusions.  These results are relevant, for instance, to the study of focal conic defects with positive Gaussian curvature~\cite{Fournier,FCD2}.

The free energy for a smectic-A liquid crystal is
\begin{equation}
F = \frac{B}{2} \int d^3x \left\{  \left[ \partial_z u - \frac{1}{2} (\mathbf{\nabla} u)^2 \right]^2 + \lambda^2 (\mathbf{\nabla} \cdot \hat{\mathbf{n}})^2 \right\}
\end{equation}
where $B$ is the bulk modulus, $\lambda$ is a length defined by the fact that $\lambda^2 B$ is the bending modulus of the smectic layers, $\hat{\mathbf{n}}= \left(\hat{\bf z} - \mathbf{\nabla}u\right)/\vert \hat{\bf z} - \mathbf{\nabla}u\vert$ is the unit normal to the smectic surface, and $u(\mathbf{x})$ is the Eulerian displacement field.  The first term accounts for the compression strain between layers while the second term arises from the bending stiffness of the layers. The
latter is manifestly rotationally invariant; the former can be shown to be rotationally invariant as well.  This nonlinearity is required and symmetry dictates the relative numerical factor of $\frac{1}{2}$ between the first- and second-degree terms.

In the limit of small strains, it is typical to retain only terms quadratic in derivatives of $u$.   This leads to a linear elasticity which misses much of the essential physics.  Instead, we make the approximation that $\partial_z u \sim (\grad u)^2 \ll 1$, where the notation $\grad$ has been used to represent the two dimensional gradient in the plane of the layers.  In the following, we will write $\partial_i u$ to represent the in-plane derivatives and continue to write $\partial_z u$ for the derivative along the layers at infinity.   In this limit,  we rewrite the free energy density as:
\begin{eqnarray}\label{eq:freeenergy}
 \label{eq:completesquare}
f& = & \frac{B}{2} \left[ \partial_z u - \frac{1}{2} \left(\grad u \right)^2 \mp \lambda \grad^2 u \right]^2 
\\&&\pm B \lambda \left[  \partial_z u\grad^2 u -\frac{1}{2}\left(\grad u \right)^2 \grad^2 u\right]\nonumber
\end{eqnarray}
We shall see that the energy density is the sum of a perfect square and a total derivative under
the appropriate conditions.  This sort of form for the free energy has been used successfully to study
the fluctuations of two-dimensional smectics~\cite{golwang} and to study shape changes in vesicles \cite{benoit}.  We use it here to study defects.

We can simplify this expression further by considering another total derivative, the Gaussian curvature $K=\frac{1}{2}\nabla\cdot\left[{\bf n}\left(\nabla\cdot{\bf n}\right) - \left({\bf n}\cdot\nabla\right){\bf n}\right]$, 
the divergence of a vector~\cite{gausscurv}.  To lowest order in the displacement field $u$, 
$ K\approx \bar K =  \frac{1}{2}\partial_i \left( \partial_i u \partial_j^2 u - \partial_j u \partial_j \partial_i u \right)$.  We decompose the vector into longitudinal and transverse components:
\begin{equation}\label{eq:Kidentity}
\partial_i u \partial_j^2 u - \partial_j u \partial_j \partial_i u  = - \epsilon_{i j} \partial_j \phi - \partial_i \psi
\end{equation}
for scalar fields $\phi$ and $\psi$.    Using this expression, we are able to rewrite the last terms in (\ref{eq:completesquare}) as:
\begin{eqnarray}\label{commonterm}
\partial_z u \grad^2 u&=& \partial_i \left(\partial_z u \partial_i u\right) - \frac{1}{2} \partial_z \left(\partial_i u\right)^2 \\
 3\left( \grad u \right)^2 \grad^2 u & = &
-4\bar K u  +  \partial_i \left(\partial_i u \partial_j u \partial_j u\right) \nonumber\\&&\qquad- 2 \partial_i \left(u\epsilon_{i j}  \partial_j \phi +u\partial_i\psi\right)
\end{eqnarray}
We see immediately that if $\bar K=0$ (as would be the case for a one-dimensional deformation such as an edge dislocation), the free energy is a perfect square plus a series of surface terms, {\sl i.e.} the 
BPS form.   Note that since $\int  d^2x \bar K$ is a constant depending on the boundary conditions (similar to the topological invariant $\int d^2x K$), a constant shift of $u$ merely shifts the zero of energy so translation invariance, $u\rightarrow u + \epsilon$, is preserved.

A result that follows immediately from this observation is that the energy of any deformation with $\bar K=0$ is bounded from below by the boundary contributions to the free energy.  This bound is saturated by certain minima of $F$, namely those for which the perfect square term vanishes and which, consequently, satisfy a first-order differential equation in contrast with the second-order equation which would follow from extremizing the free energy.  If we consider a deformation that depends only on $z$ and $x$ then $\bar K=0$ and so if
\begin{equation}\label{eq:exactequation}
\partial_z u - \frac{1}{2} \left(\partial_x u\right)^2 \mp \lambda \partial_x^2 u = 0,
\end{equation}
then the free energy will depend only on the boundary terms.
Through the Hopf-Cole transformation $S_\pm = e^{\pm u/ 2 \lambda}$ we see that $S_\pm$ satisfies the diffusion equation $\partial_z S_\pm = \lambda\partial_x^2 S_\pm$.
Solving equation (\ref{eq:exactequation}) we find that the positive branch of the theory describes deformations for $z>0$ and the negative branch describes deformations for $z<0$.  By ``gluing'' the two solutions together, we can determine the free energy merely by considering the boundary at $z=0$. 

For a single edge dislocation, the boundary conditions are $S_\pm \rightarrow 1$ as $x \rightarrow -\infty$ and $S_\pm \rightarrow e^{\pm b/4 \lambda}$ as $x \rightarrow +\infty$, where $b$ is the Burgers vector of the dislocation.  We find
\begin{equation}
S_\pm = A_\pm {\rm E}\left(\frac{x}{2\sqrt{\lambda z}}\right)  + C_\pm
\end{equation}
where ${\rm E}(x) = (\pi)^{-1/2}\int_{-\infty}^x dt \exp(-t^2)$, $C_\pm = 1$ and $A_\pm = \left(e^{b/4\lambda}-1\right)$.  It is straightforward to invert the Hopf-Cole transformation to find agreement with the edge dislocation deformation calculated in~\cite{edgedislocationtheory}.
Since boundary terms are essential to our analysis, we note that this solution can also be obtained through the boundary condition, $u_\pm(x,z=0) = \pm b \Theta (x)/2$ where $\Theta(x)$ is the step function.

The contribution to the line tension from the deformation of the layers results from a single boundary term from each region, $z>0$ and $z<0$.  This can be calculated by approximating the step function at $z=0$ with the integral of a Gaussian of width $\xi$, resulting in the expression $\tau = F/L_y = B \lambda b^2 / 3 \sqrt{\pi} \xi$.  This has the same form as the line tension of an edge dislocation in the quadratic approximation for the free energy up to a rescaling of the microscopic length $\xi$.  The reason for this is simple: were we to rewrite the energy of the linearized theory in the same form as (\ref{eq:completesquare}), we would find only one surface term.  However, in either the linear or nonlinear theories the boundary is the $xy$-plane and the free energy comes entirely from total derivatives with respect to $z$.  Thus only the (common) second term in (\ref{commonterm}) contributes to the energy and this BPS configuration of the nonlinear theory has precisely the {\emph same} energy as the same solution of the linear theory, though, as seen in experiment~\cite{edgedislocationexpt}, the actual layer deformations are quite different.   Though our solutions are absolute minima for
our imposed topology, we have not explored the possibility of modifying the defect geometry while
maintaining the same boundary conditions at infinity.  For instance, it is possible that focal conic defects may introduce additional boundaries in the bulk which could lower the energy further.  A more complete analysis will be necessary to address this issue \cite{us}.

Because $S$ satisfies a linear equation, we can superpose solutions to calculate the
interaction between two, parallel edge dislocations.   For two parallel edge dislocations at $z=0$ and positions $x_1 < x_2$ and with Burgers vectors $b_1$ and $b_2$ respectively, we have
\begin{eqnarray}
&&e^{\mathrm{sgn}(z) u(x,z)/2 \lambda}  =  1+\left(e^{b_1/4 \lambda}-1\right){\rm E}\left(\frac{x-x_1}{2\sqrt{\lambda z}}\right) \nonumber\\
&&\qquad\qquad + e^{b_1/4 \lambda} \left(e^{b_2/4 \lambda} - 1\right){\rm E}\left(\frac{x-x_2}{2\sqrt{\lambda z}}\right).
\end{eqnarray}
The layer deformation for a set of four, equally spaced parallel edge dislocations located at $z=0$ is shown in Figure~\ref{fig:layers} in comparison to the deformations derived using a linear elastic theory.  Since only one boundary condition can be specified (at $z=0$), one must explicitely check that the BPS solution satisfies the remaining boundary condition, $\partial_z u \rightarrow 0$ as $z \rightarrow \infty$ in this case.  Due to this, multiple edge dislocations solutions in different layers are not BPS configurations unless they are infinitely far apart.

The energy of a multiple edge dislocations can be determined in an analogous manner to that of a single edge dislocation.  Since the free energy is entirely determined by the same boundary term in both cases, the free energy of many edge dislocations must be equal to the free energy of those edge dislocations in the harmonic approximation.  As a result, the interaction energy between edge dislocations in the same layer must also agree.  This result is in stark contrast to the interactions of screw dislocations.  There, in the linear theory, screw defects interact exponentially, while in
the nonlinear theory they interact via a power-law repulsion~\cite{KamLub}.

\begin{figure}
	\resizebox{3.2in}{!}{\includegraphics{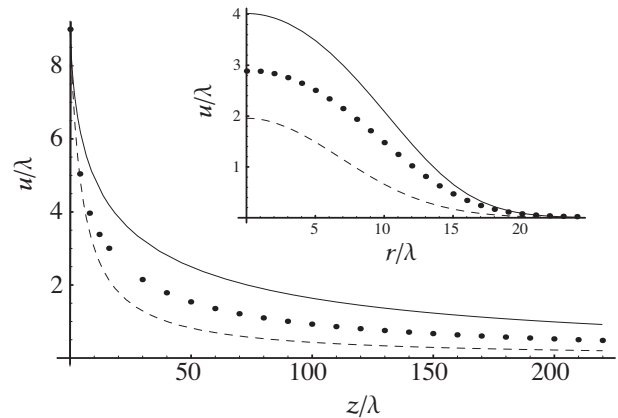}}
	\caption{\label{fig:peakdecay2} The layer displacement (points) at $r=0$ for the boundary condition of $u(r,z=0) = 9 \lambda \exp (-r^2/25 \lambda^2)$, comparing the numerical minimization with the nonlinear (solid line) and linear (dashed line) elasticity.  In the inset is a comparison of the peak shape at $z = 50 \lambda$ for all three cases.  Except for a region about $r=0$, the BPS approximation tends to be closer to the numerical minimization.}
\end{figure}

Up to this point, solving Eq. (\ref{eq:exactequation}) led to exact minima of the free energy.  We could interpret that equation as equating the strain, $u_{zz} = \partial_z u - \frac{1}{2}\left(\grad u\right)^2$, to
the bending $\lambda \bar H=\lambda \grad^2 u$ which is physically compelling: it equates the
two terms in the energy in an attempt to minimize the frustration between the two \cite{diKam}.  For more general boundary conditions ($\bar K \ne 0$), the free energy is not in the BPS form.
Nonetheless, if the additional term, $-4 B \lambda u \bar K$, is small compared to the energy scale set by the curvature, $B (\lambda \grad^2 u)^2$ (which is the only other energy scale), we can presumably treat that term as a perturbation.  In order to test this hypothesis, we have compared the layer deformation due to a Gaussian inclusion with the results from a numerical minimization using a conjugate-gradient algorithm (results shown in Figure~\ref{fig:peakdecay2}).  The layer displacement at $r=0$ falls between the BPS solution and the harmonic solution.  Though the harmonic solution is more accurate for small $r$,  overall the BPS approximant is closer to the numerical solution.  This is not surprising: because $\bar K u\gg 1$ near $r=0$ we should not expect the BPS solution to be valid there.   As we choose boundary conditions with smaller values of $\bar K$, the BPS solutions become more and more accurate.  Further study is needed to understand the precise role of the $u \bar K$ in the failure of the BPS configurations to minimize the energy~\cite{us}.  It is often the case that ``near-BPS'' solutions
are remarkably good approximants \cite{cvetic} and it appears to be true here as well.

We consider a spherical inclusion of radius $R$ located at $z=0$ which implies the boundary conditions $u_\pm(r,z=0)  =  \pm \sqrt{R^2 - r^2} \Theta(R-r)$ for $z>0$ and $z<0$ respectively, where $r = \sqrt{x^2+y^2}$, and $u_\pm(r,z)$ vanishes for $r \rightarrow \infty$ and $z \rightarrow \pm\infty$.
Solving $u_{zz}=\lambda \bar H$, we have
\begin{eqnarray}\label{eq:solution}
&&u(\mathbf{x},z) = 2 \lambda \mathrm{sgn}(z) \Bigg\{1+\frac{1}{4 \pi \lambda |z|} \\ \nonumber
&&\quad\times\int~d^2x' \exp \left[ -\frac{(\mathbf{x}-\mathbf{x}')^2}{4 \lambda |z|} \right] \left[ e^{u_{\pm}(\mathbf{x'},0)/2 \lambda}-1 \right] \Bigg\}.
\end{eqnarray}
where $\mathbf{x}$ are coordinates in the plane of the layers. 

Far from the inclusion the strain in small and  we perform a multipole expansion for $r,z\gg R$ and find that the monopole term dominates.  We have
\begin{equation}\label{eq:monopole}
u(r,z) = 2 \lambda \ln \left( 1+\frac{\beta}{z} e^{-r^2/4 \lambda z} \right),
\end{equation}
the deformation of a monopole in the region $z > 0$, where $\beta = R \left(e^{R/2 \lambda} - \frac{R}{4 \lambda} \right) + 2 \lambda \left( 1-e^{R/2 \lambda} \right)$.  Due to its exponential dependence on the radius of the inclusion, we see that for $R > 2 \lambda$, $\beta \approx R \exp (R/2 \lambda)$  becomes very large and so (\ref{eq:monopole}) differs greatly from the linear solution $u_l = \beta e^{-r^2/4\lambda z}/z$.   
Higher order moments may also be calculated explicitly for a spherical particle.  

With the success of the single monopole solution, we turn to the interaction of two inclusions that are
well separated.  We expect the deformation to be well-approximated by a superposition of BPS monopole configurations.  For two monopoles of size $\beta_1$ and $\beta_2$ separated by a distance $2 x_0$ and both located at $z=0$, the deformation profile can be found by applying superposition.
By expanding in powers of $\beta_i$, we can estimate the interaction energy.  The lowest order term is, in principle, quadratic in $\beta_i$ and identical to the interaction in the harmonic approximation and higher order terms arise from the
nonlinearity of the elasticity:
\begin{eqnarray}\label{eq:interaction}
F_I & = &-\frac{24 \pi \beta_1 \beta_2 \left( \beta_1+\beta_2 \right)}{x_0^6}\\
& & +\frac{20 \pi \beta_1 \beta_2 \left(512 \beta_1^2 + 245 \beta_1 \beta_2 + 512 \beta_2^2\right)}{81 x_0^8} + \cdots.\nonumber
\end{eqnarray}
Notice that the term quadratic in the monopole moments vanishes and thus, as in the case of screw dislocations \cite{KamLub} the interaction energy is essentially nonlinear.  This expression for the energy compares reasonably with our numerical estimates.  

We have introduced a method to calculate deformations of a smectic liquid crystal in a way that takes into account nonlinearities in the elasticity, as well as their energies.  While the method is asymptotically exact for the one-dimensional deformations, we find that it produces deformations which have lower free energies than those of the usual quadratic approximation.  This allows us to calculate the asymptotic interaction energy between two large smectic inclusions.  In future work we will study the implications of nonlinearities of dislocations with both screw and edge components, and the nonlinear elasticity of the full smectic free energy.

\begin{acknowledgements}
We thank P. Pincus for continued guidance throughout the course of this work, M. Cveti\v c and P. Sens for interesting discussions, and R. Sedgewick for countless useful converstions on numerical methods.  CDS was supported by the MRL Program of the National Science Foundation under Award No. DMR00-80034.  RDK was supported through NSF Grant DMR01-29804 and a gift from L.J. Bernstein.
\end{acknowledgements}

\end{document}